# Enhancing Live Broadcast Engagement: A Multi-modal Approach to Short Video Recommendations Using MMGCN and User Preferences

**Saeid Aghasoleymani Najafabadi[1,*], Elaheh Nabavi Nia[2]**

[1] Faculty of Industrial Engineering, Urmia University of Technology, Urmia, Iran
[2] Harrisburg University of Science and Technology, McLean, VA, USA
*Corresponding author: s.aghasoleimani@ine.uut.ac.ir

**Abstract**

The purpose of this paper is to explore a multi-modal approach to enhancing live broadcast engagement by developing a short video recommendation system that incorporates Multi-modal Graph Convolutional Networks (MMGCN) with user preferences. To provide personalized recommendations tailored to individual interests, the proposed system considers user interaction data, video content features, and contextual information. With the aid of a hybrid approach combining collaborative filtering and content-based filtering techniques, the system can capture nuanced relationships between users, video attributes, and engagement patterns. Three datasets are used to evaluate the effectiveness of the system: Kwai, TikTok, and MovieLens. Compared to baseline models, such as DeepFM, Wide & Deep, LightGBM, and XGBoost, the proposed MMGCN-based model shows superior performance. A notable feature of the proposed model is that it outperforms all baseline methods in capturing diverse user preferences and making accurate, personalized recommendations, resulting in a Kwai F1 score of 0.574, a Tiktok F1 score of 0.506, and a MovieLens F1 score of 0.197. We emphasize the importance of multi-modal integration and user-centric approaches in advancing recommender systems, emphasizing the role they play in enhancing content discovery and audience interaction on live broadcast platforms.

## 1. Introduction

Digital experiences have been revolutionized by rapid developments in internet technology and online platforms in recent years, making brief video streaming and live broadcasting an integral part of contemporary life. Especially for sites such as Netflix, Amazon, and TikTok, where personalized content distribution fuels customer satisfaction and engagement, recommender systems (RS) have become essential for today's digital age [1-3]. Through the navigation of the vast amount of information available, these systems aim to provide personalized recommendations to consumers based on their interactions and interests [4]. Although traditional recommendation algorithms are effective, they often fail to engage users fully because they are unable to capture intricate relationships between items and users and strike a balance between customization and discovery [5,6]. The objective of this study is to overcome these constraints by introducing a new short video recommendation system that incorporates user preferences with Multi-modal Graph Convolutional Networks (MMGCN) to improve live broadcast engagement. This system utilizes a variety of data sources, including contextual information, video content properties, and patterns of user interaction, to deliver highly customized suggestions.

By combining collaborative filtering algorithms with content-based filtering algorithms, the proposed method generates personalized suggestions that correspond to individual preferences. For this system to be successful, graph neural networks are used, which are excellent at simulating intricate interactions through efficient information flow and aggregation across graph nodes. A multi-modal graph structure is employed to represent user preferences and video characteristics to enhance the precision of feature extraction and rating predictions. Moreover, the integration of attention networks ensures that the fluctuating significance of distinct neighbors is dynamically recorded, allowing for accurate and pertinent suggestions. Inspired by advancements in interdisciplinary education that employ simulation-based models to explore complex systems such as the vestibular system in human physiology [5] this study adopts a multi-modal, graph-based approach to simulate user-video interactions for more adaptive and personalized video recommendations. This study enhances recommendation accuracy by leveraging MMGCN, a multi-modal database that incorporates user-video interactions. Several experiments have demonstrated that the system can adapt to a wide range of user preferences and provide recommendations for brief videos that will increase user satisfaction and engagement. Multi-modal integration and user-centric strategies are important in promoting content discovery and increasing audience engagement. This study advances the field of recommender systems by emphasizing the benefits of multi-modal integration and user-centric strategies.

## 2. Literature review

Recommendations can take many forms, including content recommendations, collaborative filtering, demographic recommendations, etc. With these recommendation approaches and various machine learning techniques, we can improve movie recommendations. The recommendation system can use different methods and techniques in machine learning to improve user recommendations. Table 1 summarizes the literature review in personalized recommendation systems, including key methodologies, strengths, and limitations.

**Table 1:** Literature review summary

| Author Name | Year | Method | Aim | Key Results |
|---|---|---|---|---|
| Mo et al. [64] | 2025 | A fusion model combining modality-behavior alignment with graph contrastive learning | To enhance recommendation performance under sparse user-item interactions in multimodal settings | FGCM improved accuracy and robustness by filtering interaction noise and effectively fusing multimodal signals, outperforming several state-of-the-art approaches on three datasets |
| Wang et al. [63] | 2025 | Multi-modal Negative Sampling with User Interest (MNS-UI) | To capture latent semantic relations and preserve user interest during negative sampling in multimodal RSs | MNS-UI improved Recall@20 by 2.4%, NDCG@20 by 6.7%, and reduced training epochs by 87.3% |
| Zhao et al. [60] | 2025 | METRIC framework combining Multiple | To improve personalized | METRIC captures fine-grained user interests, |
| | | Preference Modelling (MPM) | recommendations by refining multimodal relationships and user preference modeling | enhances semantic understanding, and performs well across three benchmarks |
| Chen et al. [59] | 2025 | Multi-Modal Adversarial Method (MMAM) | To provide reliable personalized recommendations under missing modality scenarios | MMAM improves prediction accuracy and handles missing modalities better than existing methods |
| Guo et al. [58] | 2025 | Multi-Modal Hypergraph Contrastive Learning (MMHCL) combining user-user and item-item relations | To address data sparsity and cold-start issues using multimodal relationships | MMHCL produced denser embeddings and better feature distinction, improving recommendation accuracy |
| Farhadi Nia [62] | 2025 | LLMs (e.g., ChatGPT) applied to dental diagnostics using NLP and datasets | To improve diagnostic accuracy and communication in dentistry | Showed ChatGPT's potential in clinical support and oral health innovation |
| Prasad et al. [6] | 2024 | Machine learning, Combined similarity-based filtering | To create a recommendation system using different filtering techniques to address information overload | Better recommendations than traditional methods; also addresses the cold start problem |
| Tian [7] | 2024 | Content-based Filtering (CBF) | To improve movie recommendation systems through content-based filtering | Identified CBF challenges and suggested personalized delivery improvements |
| Norcéide et al. [15] | 2024 | Integration of AR with Neuromorphic Vision Sensors (NVS) | To create an accurate and power-efficient real-time object tracking system in AR | AR-NVS reduced data load, improved lighting adaptability, and enhanced AR interaction |
| Chen et al. [70] | 2024 | Multifactorial Modality Fusion Network (MMFN) using GNNs | To enhance modality fusion and training efficiency | MMFN outperformed baselines by up to 24.51% in Recall@20, showing high training efficiency |
| Abbasalizadeh et al. [61] | 2024 | LGLS algorithm | A novel technique for optimizing wireless network link scheduling | The LGLS algorithm is a robust and efficient solution for future wireless network optimization |
| Farhadi Nia et al. [57] | 2024 | Quality-Rate analysis using convex hull modeling on H.264, H.265, VP9 | To reduce bitrate while maintaining quality in adaptive streaming | Accurately predicted PSNR and VMAF with convex hulls; supported efficient codec selection |
| Widayanti et al. [8] | 2023 | Collaboration-based filtration combined | To enhance recommendations by integrating CF and CBF | Improved diversity and precision; overcame CF and CBF limitations individually |
| | | with content-based filtration | | |
| Parthasarathy & Sathiya [9] | 2023 | Hybrid of collaboration and content-based filtering | To develop an optimized hybrid recommender based on user preferences | New hybrid system outperforms in accuracy, FDR, MAE |
| Sharma et al. [10] | 2023 | Integrated approach using content-based and collaborative learning | To build a movie recommender using ML for better user experience | Reduced human input; used ML to improve suggestions |
| Kumar et al. [11] | 2023 | Content-based Recommender System | To propose an ML-based movie recommender | Applied specific algorithms and demonstrated ML effectiveness |
| Muhammad & Rosadi [12] | 2023 | Collaborative Filtering (User-based and Item-based) | To compare item-based and user-based CF | IBCF outperformed UBCF in prediction accuracy |
| Vuong Nguyen et al. [13] | 2023 | Combining CF and word embedding | To extend CF recommendations using word embeddings for movies | Improved similarity and filtering accuracy using plot embedding |
| Ahmadi et al. [65] | 2023 | Comparison of U-Net and pretrained Segment Anything Model (SAM) | To develop and evaluate deep learning models for accurately segmenting medical image | U-Net outperformed SAM in detecting and segmenting tumor regions, especially in complex cases; SAM showed limitations with malignant tumors and weak boundaries |
| Afoudi et al. [14] | 2021 | Hybrid of content-based, collaborative filtering, and neural networks | To introduce a hybrid recommender for improved performance | Outperformed traditional methods in accuracy and precision |
| Nallamala et al. [16] | 2020 | Filtering algorithm based on collaborative content | To analyze CF, CBF, and hybrid algorithms | Reviewed strengths and weaknesses of major filtering approaches |

## 2.1. Live Streaming Gifting Recommendation

There are currently no live streaming gifting recommendation systems that consider the entire live room as a recommendation target, and those that do use categorical data to model the interaction between streamers and viewers. MARS [11] proposes a novel recommendation scenario called Multi-Stream Party (MSP), which involves designing two-phase methods to optimize both donation reciprocity and personal satisfaction within the MSP. The LSEC-GNN model [55] models the live stream e-commerce scenario using GNN and leverages information about interactions between streamers, users, and products. Research has neglected the fact that dramatic content changes can occur even within the same living room, so it's essential to take full advantage of the multi-modal feature. To address this issue, MTA proposes a novel orthogonal projection model that captures cross-modal information interaction in real-time. However, MTA models users' interests as time series predictions instead of individualized gifts. To conclude, existing methods for predicting live-streamed gifting still have room for improvement

## 2.2 Personalized Recommendation

Deep neural networks are widely used to make personalized recommendations. The DIN model incorporates attention mechanisms to model users' diverse interests. According to SIM, an online two-stage retrieval method based on the features of the current candidate item models relevant behaviors from a user's long-term history. However, live-streaming gifting scenarios pose challenges due to sparsity issues in streamers and user interactions. Multiple recent studies have introduced multimodal features to enhance graph node embedding using GNNs.

# 3. Recommender System

In a recommender system, user information is discovered and analyzed to generate appropriate suggestions for every user. Recommender systems have been defined in a variety of ways [17]. The broadest and most concise definition is Ting-peng Liang's 2007 definition, which refers to RS (Recommender Systems) as a subset of DSS (Decision Support Systems) and defines it as an information system that identifies past behaviors and provides recommendations for the future. A recommendation system identifies and recommends the most suitable and closest product to a user's taste (using data from their behavior and that of similar users). These systems mimic what we do every day, when we ask for opinions about our choices from people with similar tastes to ours [18 ,56]. The recommendations provided by recommended systems can generally lead to two outcomes:

- They assist users in deciding (for example, choosing the best option among several available ones).
- They increase the user's awareness in their area of interest (for example, by introducing the user to new items or subjects they were previously unaware of). Recommender systems are beneficial for both parties in an interaction (commercial or non-commercial) and provide advantages. For instance, in commercial interaction, customers find these systems useful as they facilitate and expedite the search process with a large volume of information; sellers can increase customer satisfaction and boost their sales with the help of these systems.

## 3.1. Advantages and Advances

Many web users have difficulty making decisions and selecting information, data, or goods due to the large and growing volume of information available on the internet. Researchers sought a solution to this fundamental problem of the modern age, known as data overflow, because of this issue itself. There have been two approaches proposed so far to address this issue [19]. In the first approach, two concepts were utilized: information retrieval and information filtering. The primary limitation of these two concepts in providing recommendations is that, unlike human recommenders (such as friends, family members, etc.), these methods cannot discern and distinguish between high-quality and low-quality items in making recommendations for a topic or product [20]. A second approach, known as the recommender system, was developed in response to this problem. As a result of these new systems, the issues encountered by the previous systems have been resolved.

## 3.2. Applications for Recommender Systems

Across a wide range of industries, reciprocal systems demonstrate their versatility and value. By providing consumers with a range of products and services, e-commerce tailors the shopping experience to their individual needs [21]. Incorporated intranets can benefit from these systems by identifying experts in specific fields or individuals who have valuable experience in addressing specific challenges, making them particularly useful within organizational settings. To facilitate the discovery of books, articles, and other resources, digital libraries utilize recommender systems. Medical recommender systems assist in matching patients with doctors who will treat their specific conditions based on factors such as location, type of illness, and availability, as well as in selecting appropriate medications [22-24]. Additionally, in the field of Customer Relationship Management (CRM), these systems offer strategies aimed at bridging the gap between producers and consumers within the supply chain, to resolve issues and enhance customer satisfaction. Recommender systems offer an invaluable tool to navigate the vast number of options available in a variety of domains, making them more accessible and tailored to individual needs and preferences.

## 4- Comparison of Recommender Systems and Classic Decision Support Systems

There are many similarities between these two systems; however, there are also several important differences, the most significant of which is that in DSS (Decision Support Systems), end users are senior or middle managers of an organization, whereas recommender systems are not limited to a specific level and are generally applicable to all levels. While there are many similarities between these two systems, the major difference is that recommender systems are technically considered to be a subset of decision support systems. These two systems are information systems that provide users with knowledge, databases, user interfaces, etc., that assist them in making decisions.

## 5- Recommender System Types

Content-based systems can be divided into two types: the knowledge-based approach and the third approach. Hybrid RS is a fourth category recognized by collaborative filtering (CF). A collaborative filtering algorithm is one way to implement recommender systems [25]. As a result, users' opinions and ratings are used to provide recommendations instead of the content of items. It is recommended that a user purchase items that are like items the active user (the user we recommend) has expressed interest in in the content-based method. However, in CF, the recommended items are selected based on whether the active user has been satisfied with them in the past. As a result, content-based methods aim to find similarities between items, while user-based methods aim to find similarities between users. In CF, recommendations are based on behavioral similarities between users, not on the similarities between recommended goods and products the active user is interested in. Known as knowledge-based systems, the third type is based on information.

These systems provide recommendations based on the understanding they have gained of the customer's needs and the features of the goods. In other words, in these types of recommender systems, the raw material used to generate a list of recommendations is the system's knowledge about the customer and the product [26,27,28]. Knowledge-based systems can be analyzed using a variety of methods, including genetic algorithms, fuzzy logic, neural networks, etc. These systems can also use decision trees, case-based reasoning, etc. Case-based reasoning or CBR is one of the most common methods of knowledge analysis in knowledge-based recommender systems. The fourth type of system is a hybrid one. This type of system is often designed by combining two or more of the three types mentioned above for two main reasons: to improve system performance and to reduce the weaknesses that these systems have when used separately. The CF method is usually present in all three existing methods (CF, CB, and KB) [27].

## 6. Classification of Recommender Systems

Traditional and modern recommender systems generally fall into two categories.

## *6.1 Traditional Recommenders*

### *6.1.1 Content-based Filtering*

A user's rating of an item is considered to identify the items that are of interest to them based on their rating. Following this, the system recommends similar items to the user. Among the techniques used in this method are Bayesian classifiers, clustering, decision trees, and artificial neural networks [28].

### *6.1.2 Collaborative (Social) Filtering Systems*

Similar users are identified by these systems and then items that they have liked are recommended. A similarity function [29] is used to identify similar users.

### *6.1.3. User-User Collaborative Filtering*

A group of users who have rated similar items is formed. Based on the ratings they give to other items, predictions are made. Cosine similarity, Pearson correlation, etc., are used in a similar manner. For prediction, we select the top users with the highest similarity (weight) to the active user, and we assign each user a weight based on that similarity. Neighborhood-based filtering is also known as this method. Among the disadvantages are computational complexity and the need to search for similar users [30]. According to this approach, ratings are not taken into account when clustering items, but rather item features are taken into account. A second method considers each item as a node in a Bayesian network. The system does not account for the fact that a user might belong to more than one cluster. An individual may be interested in both programming books and novels, for example. The results are improved when this method is combined with neighborhood-based methods.

## *6.2. Modern Recommender Methods*

### *6.2.1 Context-Aware Methods*

Generally, text represents a series of information about the environment (e.g., what should the system be like? A film recommender system, article recommender system, etc.) and users, considering the temporal and spatial situation for recommendations. Methods consider users who have viewed a particular item at a certain time of day or from a certain location [31]. This is done through user surveys, data mining methods, and the use of GPS to monitor the time, date, and location of individuals.

### *4.2.2 Semantic or Knowledge-based Methods*

Recommendations are made based on the system's perception of users and item features. This method uses classification algorithms like genetic, fuzzy, and neural networks, and CBR (Case-Based Reasoning) [32]. An example of semantic-based systems is the Wikipedia database system, which operates based on the classification of topics and users. Hierarchical or three diagrams for representing movies can be used in a movie recommender system. Ontologies are also used to express relationships between items and users [33].

# 7. Terms and Definitions in Evaluating Recommender Systems

A recommender system can be understood by considering the following four fundamental concepts. The user for whom the current recommendation is being prepared and processed is referred to as the active user or the target user in recommender systems. As part of their algorithms, these systems utilize a matrix known as the ratings matrix; other common terms for this matrix include Rating Database and Preference Database [34]. Whenever a user accepts a suggestion offered by a recommender system, the term "consume" is used. Therefore, when a user accepts a recommendation made by the system, it is said that the recommendation has been consumed. The acceptance of the recommended book can take many forms, including the purchase of the book, browsing the recommended website, or consulting the recommended service provider. A ratings matrix has the structure of a row for each user and a column for each item. We will now introduce the concept of a utility function, which will be used to present a general mathematical model of recommender systems [35]. This mapping can be used to model and equate a recommender system:

$$u: C \times S \rightarrow R \quad (1)$$

Assume that C represents the set of all users and S represents the set of available items. The function that calculates the usefulness and appropriateness of item c for user s is denoted by u, where Riis a completely ordered set (based on importance). A set of attributes can be assigned to each element of S. An example would be the title of a movie, the director, the duration of the movie, and the production date. Similarly, the elements of set C can also be recorded based on characteristics like age, gender, etc. It should be noted that is not defined over the entire space of the initial set S×C; hence, it needs to be extrapolated [36].

## 8. Systems for recommending products and services: challenges and issues

The "Cold Start" problem in recommender systems arises when new users, who have minimal or no interaction history within the system, struggle to receive personalized recommendations due to their empty or near-empty profiles [37]. To mitigate this challenge, content-based filtering or hybrid methods are often deployed, leveraging available user profile information or item attributes to generate relevant recommendations. On the other hand, for users with a substantial interaction history, collaborative filtering techniques are more commonly applied, capitalizing on the wealth of rating data to predict user preferences accurately. Furthermore, the issue of "Trust" becomes pertinent as the system's confidence in recommendations can vary based on the longevity of a user's history with the system. Users with a longer history of interactions are typically deemed more credible, suggesting that recommendations for these users can be made with higher confidence [38,66]. Addressing this, systems can prioritize users based on the length of their interaction history, thereby enhancing the overall trustworthiness and reliability of the recommender system's outputs [67].

## 9. Modeling the performance and impact of recommender systems

After gaining a comprehensive grasp of the many kinds of recommender systems and how they operate, we now focus on evaluating them. The main goal of current recommender systems is to direct users toward desirable, interesting, and helpful goods. Consequently, evaluating a recommender system entails determining how effectively it accomplishes this goal [39]. Even though these systems have developed and expanded over time, assessing their success indicators is still difficult [68]. "Just as recommender systems rely on accurate evaluation metrics to ensure reliable suggestions, privacy-aware detection in large language models also requires robust modeling to balance performance with user trust [40]. the majority of empirical assessments of recommender systems over the years have been on determining how well the system predicts user preferences [41].Generally speaking, these metrics are inconsistent with the overall objective of recommender systems. Moreover, the capabilities of recommender systems are limited by certain metrics.

There are now several ambiguous, indeterminate aims that are expressed informally using a range of metric systems in the absence of a shared target [69]. Researchers in this discipline have said the following: The large number of published measures available for evaluating the accuracy of recommender systems makes it difficult to select one that covers them all [41]. Without standardization, collaborative filtering-based recommender systems would struggle to advance. Because researchers may add new measurements while assessing their systems, it is impossible to compare study outcomes across publications without standardized metrics. Thus, compiling the diverse publications regarding recommender system algorithms into a cohesive scientific collection will be challenging. There are also specific limitations to these metrics. For quantitative comparison, researchers must first choose one or more metrics. It is difficult for researchers to choose a metric because they face a variety of issues

- Measures the system's efficiency in relation to the user tasks it was designed to handle?
- Can this metric be compared to other published results in this field?
- What assumptions underlie a metric?
- Can the metric detect real differences if it is sensitive enough?
- In order for a difference between two metric values to be statistically significant, what magnitude of difference must there be?

It has not yet been possible to provide a complete answer to these questions in the published literature. Even with these issues, we must not lose sight of the ultimate goal of any recommender system metric. In order to be a good metric, it must be able to determine the "good behavior" of a recommendation system and produce similar results whenever these systems are presented in similar circumstances (for example, using a dataset of user interactions). Furthermore, "good behavior" must be clearly defined. As a general recommendation system, it should strive to guide users to useful, interesting, and desirable things and objects [42]. Towards this goal, two distinct tasks must be completed: (a) Generating user-acceptable recommendations. Filtering items that are useful and interesting. The first task is accomplished through an external, interactive behavior presented directly to the user by each recommender system. Second, finding good items requires more internal behavior and less interactive interaction. To date, a plethora of metrics have been published, but defining these two tasks as a general objective has proven extremely challenging. Additionally, specific research may be more focused on the second objective, often ignoring the first. Our discussions will ultimately lead to the development of a new standard that fulfills both of these objectives.

## *9.1. Evaluating Recommender Systems*

Data overload is mitigated using recommendation systems that use statistical techniques and knowledge discovery methods to provide users with product recommendations. Assessing the quality of recommender systems has become crucial in selecting the best learning algorithms [43]. A standardized and universally accepted method is needed to compare different recommender systems accurately. Currently, however, the evaluation of these systems often utilizes inconsistent methods, primarily due to the absence of a common framework that is broad enough to encompass a wide range of recommender systems to date and specific enough to yield consistent and reliable results. This document introduces and explains the existing metrics for evaluating recommender systems and ultimately proposes a framework that extracts the essential features of recommender systems. In this framework, the primary focus of each recommender system is considered its core objective [44]. Different modalities have semantic gaps. Although the micro-videos in Figure 1 appear visually similar, the textural representations are dissimilar due to the topic words [1]. The modeling of item representations would be misled if such modality differences were ignored.

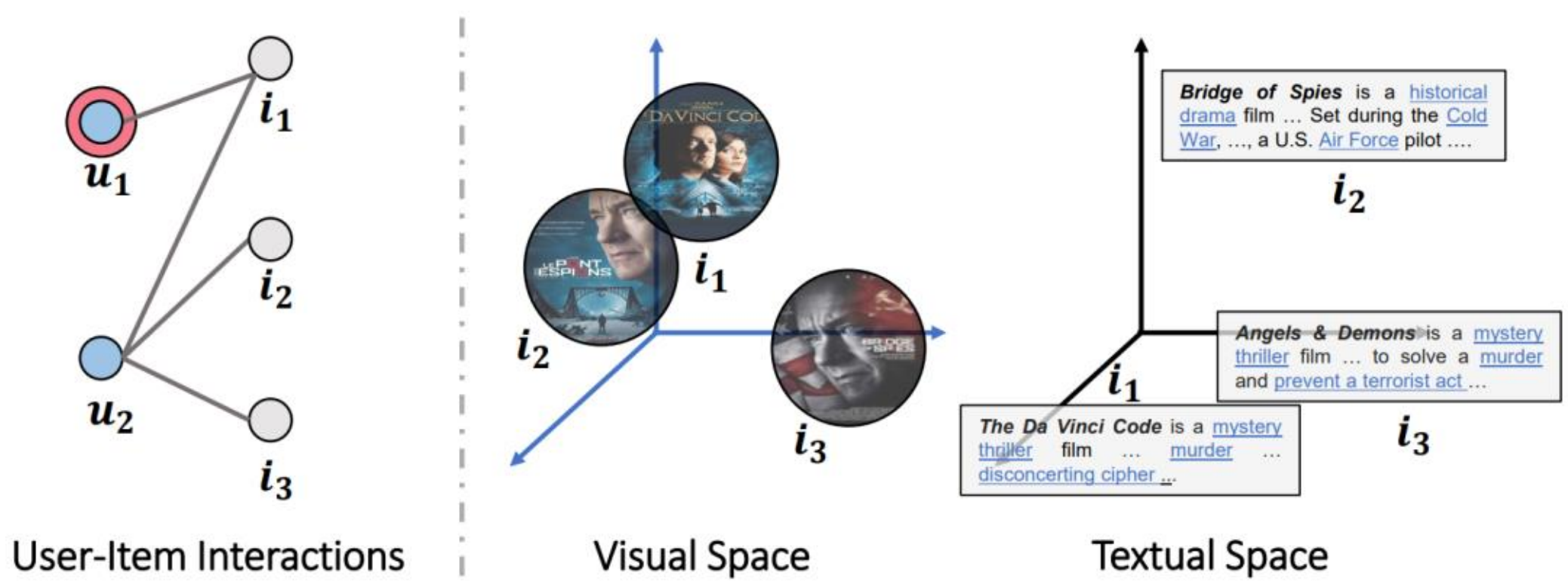

**Figure 1:** User preferences based on modals

Recommended system applications have two fundamental goals: they must (a) determine which items (from among many) should be shown to the user, and (b) determine when and how these recommendations should be presented. A new metric emerges naturally from applying this framework, as demonstrated in this document. A single metric may eventually be capable of evaluating a wide range of recommender systems by comparing the properties of this new metric with traditional metrics. The web is a vast, diverse, and dynamic environment where numerous users publish their documents. It grows through a chaotic and decentralized process, leading to a vast volume of interconnected documents without any logical organization. Given the massive amount of information on the web, managing it with traditional tools is nearly impossible, necessitating new tools and methods for its management. To address this issue, web personalization has become a popular phenomenon for customizing web environments. The goal of personalization systems is to meet users' needs without them explicitly stating or demonstrating those needs. Web personalization is a set of operations that organizes the web experience for a specific user or a group of users. As mentioned, the overwhelming and ever-growing volume of information on the web and internet has complicated the decision-making process for many web users regarding selecting necessary information, data, or goods. This challenge motivated researchers to find a solution to confront this fundamental problem of the new age known as data overflow [45]. To date, two approaches have been proposed, the first being the use of information retrieval and information filtering. Unlike human recommenders (such as friends or family members), these methods cannot distinguish between high-quality and low-quality items when making recommendations for a topic or product. This issue led to the emergence of the second approach, the recommender system, which solves the problems of the initial approach.

Recommender systems are recognized as a type of decision support or decision assistant systems that emerged as an independent research field in the 1990s, serving as one of the solutions to the information overload issue on the web. Although recommender systems have a wide range of applications, their use in e-commerce is particularly notable. On one hand, e-commerce businesses need these systems to attract more customers in a highly competitive market, and on the other hand, customers need assistance navigating the vast and growing range of products and related information to make the best choices among numerous options. Recommender systems are the primary technology for personalized information retrieval, operating in two main ways: (1) predicting whether a particular user (customer) will like a specific product, and (2) predicting the category of products likely to appeal to a user. In this way, recommender systems satisfy customer needs while also boosting sales [46].

## *9.2. Recommender System Key Performance Indicators*

Every time a recommender system is evaluated, a particular measure is applied based on a set of presumptions. Therefore, this part examines and analyzes the most popular metrics, identifies underlying assumptions, and considers the most important ones. The first step in achieving this is to outline the widely used framework in this field for setting up the basic recommendation process [47-50]. A recommender system exists in the existing structure, but it is integrated with another system that has many things that need to be suggested. To begin the suggestion process, individuals must rate the goods. This grade is available in many recommender systems. In other cases, implicit ratings are derived from other users' interactions.

After the recommender system receives a sufficient number of ratings, the procedure can begin. The recommender chooses N<=I items for each batch of recommendations, which are then shown to the intended user. Furthermore, some recommender systems score a small number of items to show the user in a well-structured list. Items near the top of this list are likely to be reviewed by the user in the following step [48]. Similar to recommender systems, where evaluation relies on ranking metrics and adaptive selection, scheduling in mmWave IAB networks also requires intelligent optimization to balance resources under dynamic conditions [51]. Finally, to evaluate the performance of the recommender system, for each object shown to a specific user, we must measure the desirability and closeness of the shown item relative to the user's preferences. Furthermore, regarding the ranked list, we also need to account for and interpret the placement of each recommended item within this list.

## *9.3. Preliminary Considerations*

The degree to which projections match customers' real preferences is usually gauged through a numerical representation. In order to maintain clarity, we will use the same notation throughout this section. A recommender system forecasts P(u,i) for a given user u and item i, whereas the user's actual preference value is p(u,i). The function p(u, i) cannot be determined with perfect accuracy. Thus, users' prior evaluations are commonly used to calculate the value of this function. Evaluations can be explicit or implicit, as was already established. In some cases, both functions p(u, i) and P(u, i) return 1 as a result of the comparison, showing whether the object i is useful or useless for the user. Due to these special circumstances, binary functions are called p and P [49-52].

## 10. Advantages of Using Recommender Systems

The utilization of recommender systems brings numerous benefits, most notably their operation on live and real-time user data, which ensures that their suggestions are grounded in actual user behaviors and preferences rather than mere speculation. This aspect is crucial for delivering relevant and personalized recommendations. Additionally, these systems excel in uncovering new domains, offering users access to content such as movies, images, and articles that they might not have discovered on their own. This capability significantly enhances the user experience by broadening their horizons and saving time that would otherwise be spent on aimless searching [53]. The personalization aspect of recommender systems mirrors the trust we place in recommendations from friends and family, as these systems make recommendations based on a deep understanding of individual user profiles, thus providing more targeted and meaningful suggestions. Moreover, by harnessing the power of collective intelligence, users are kept abreast of the latest trends and information in their areas of interest, ensuring they remain updated without substantial effort. Lastly, recommender systems offer an organizational advantage by reducing maintenance costs; since users directly contribute to and shape the content based on their preferences, the burden of managing and updating website information is significantly lessened, making it a cost-effective solution compared to traditional web management approaches [54].

## 11. Results and Discussion

### *11.1. Data collection*

The recommender system developed in this research utilizes the "MovieLens 20M ratings" dataset, which contains 20 million ratings for 27,000 unique movies provided by 138,000 users. The data spans a timeline from January 9, 1995, to March 31, 2015, thereby excluding films released after the latter date. Ratings range from 0.5 to 5.0, with each user having rated at least 20 movies. For this study, we primarily focus on two data files: "Ratings" and "Movies." These files are processed using Python's "pandas" library in the "DataFrame" structure, with visual representations of these DataFrames provided for clarity.

A defining characteristic of this dataset is the high level of sparsity in the ratings data. This sparsity arises from the extensive collection of movies compared to the relatively limited number of ratings provided by individual users. Consequently, the hypothetical user-movie matrix, consisting of 138,000 user rows and 27,000 movie columns, is predominantly filled with "NaN" values, representing missing ratings. To address this challenge and efficiently process the sparse data, we leverage the "sparse matrices" functionality provided by the SciPy library (see Table 2 to 4).

**Table 2:** Summary of dataset

| | Items | Count |
|---|---|---|
| 1 | Number of users | 610 |
| 2 | Number of ratings record | 100836 |
| 3 | Number of movies | 193609 |

**Table 3:** Head of the rating database

| | userID | MovieID | rating |
|---|---|---|---|
| 0 | 1 | 1 | 4 |
| 1 | 1 | 3 | 4 |
| 2 | 1 | 6 | 5 |
| 3 | 2 | 47 | 5 |

**Table 4:** Head of the movies database

| MovieID | MovieID |
|---|---|
| 1 | Toy Story (1995) |
| 2 | Jumanji (1995) |
| 3 | Grumpier Old Men (1995) |
| 4 | Waiting to Exhale (1995) |
| 5 | Father of the Bride Part II (1995) |

In order to develop and evaluate video recommendation systems, TikTok and Kwai, two widely used short-video platforms, provide extensive datasets. In the Kwai dataset, there are 1,664,305 interactions among 22,611 users and 329,510 items. There are 2,048 visual features and 100 textual features, but no acoustic features. It is ideal for exploring content-based recommendations and user preference modeling in localized and community-driven scenarios because it emphasizes large-scale interactions between users and items.

### *11.2. Data Preprocessing*

Initially, we import the pandas library, which provides us with the tools necessary to manipulate and analyze our data. We have two primary components in our dataset: ratings.csv and movies.csv. In the former, users rate various movies, while in the latter, a list of the movies included in the dataset is presented. We ensure that only movies with more than 10 ratings and users who have rated more than 10 movies are included in our database through careful manipulation. In order to maintain a high-quality dataset, we must filter out outliers or sparse data that might skew our analysis.

### *11.3. Construction of the Rating Matrix*

As part of the development of our movie recommender system, the construction of the rating matrix is a crucial step. Essentially a two-dimensional array, this matrix is used to uncover patterns in user preferences and to provide personalized movie recommendations. This matrix reflects the complex interactions between users and movies, captured by ratings that range from 0.5 to 5.0. It provides insight into the nature of our dataset and the collaborative filtering approach at its core. As a result of this method, our ratings data is transformed into a matrix, in which each row represents a user, each column represents a movie, and each cell represents the rating given by that user. The absence of a rating is indicated by the use of NaN, which indicates that the movie has not been rated by a user. In constructing the rating matrix, our initial step involves filtering the data to ensure its quality and relevance. We start by removing movies with fewer than 10 ratings and users who have rated fewer than 10 movies. This filtration process is critical for two reasons. First, it mitigates the impact of noise in the data, such as outlier ratings or movies with minimal information that could distort our analysis. Second, it helps to address the challenge of sparsity in the matrix. Even after filtering, the matrix remains sparse, as most users have only rated a small fraction of the entire movie catalog.

However, by focusing on more actively rated movies and users, we improve the density of meaningful data points in our matrix. Upon constructing the filtered rating matrix, we gain valuable insights into the user-movie interactions within our dataset. The matrix's sparsity is a significant challenge in collaborative filtering systems, underscoring the importance of sophisticated techniques to extract useful patterns from the available data. To illustrate, our resulting matrix spans 608 rows (users) and 2121 columns (movies), with most cells being NaN values. This sparsity highlights the selective nature of movie ratings: users tend to rate only a subset of movies that they have watched, leaving a large portion of the matrix unfilled. To further explore the rating matrix, we examine the distribution of ratings across movies and users. For instance, some movies, such as "Forrest Gump" and "Shawshank Redemption," receive a high number of ratings, indicating their popularity and widespread appeal. Conversely, other movies might only have the minimum threshold of ratings, suggesting a more niche or specialized audience. Similarly, the distribution of ratings per user varies significantly, with some users being very active in rating movies while others contribute minimally beyond the threshold.

## 12. Model framework

Our proposed framework is composed of three key components, as shown in Figure 2 [1], the aggregation layer, the combination layer, and the prediction layer. Using multiple layers of aggregation and combination, the framework effectively captures information exchange between users and items in each mode of representation learning. Last but not least, the multimodal representations are incorporated into the prediction layer to estimate how users interact with micro-videos. Each component is explained in detail in the following sections.

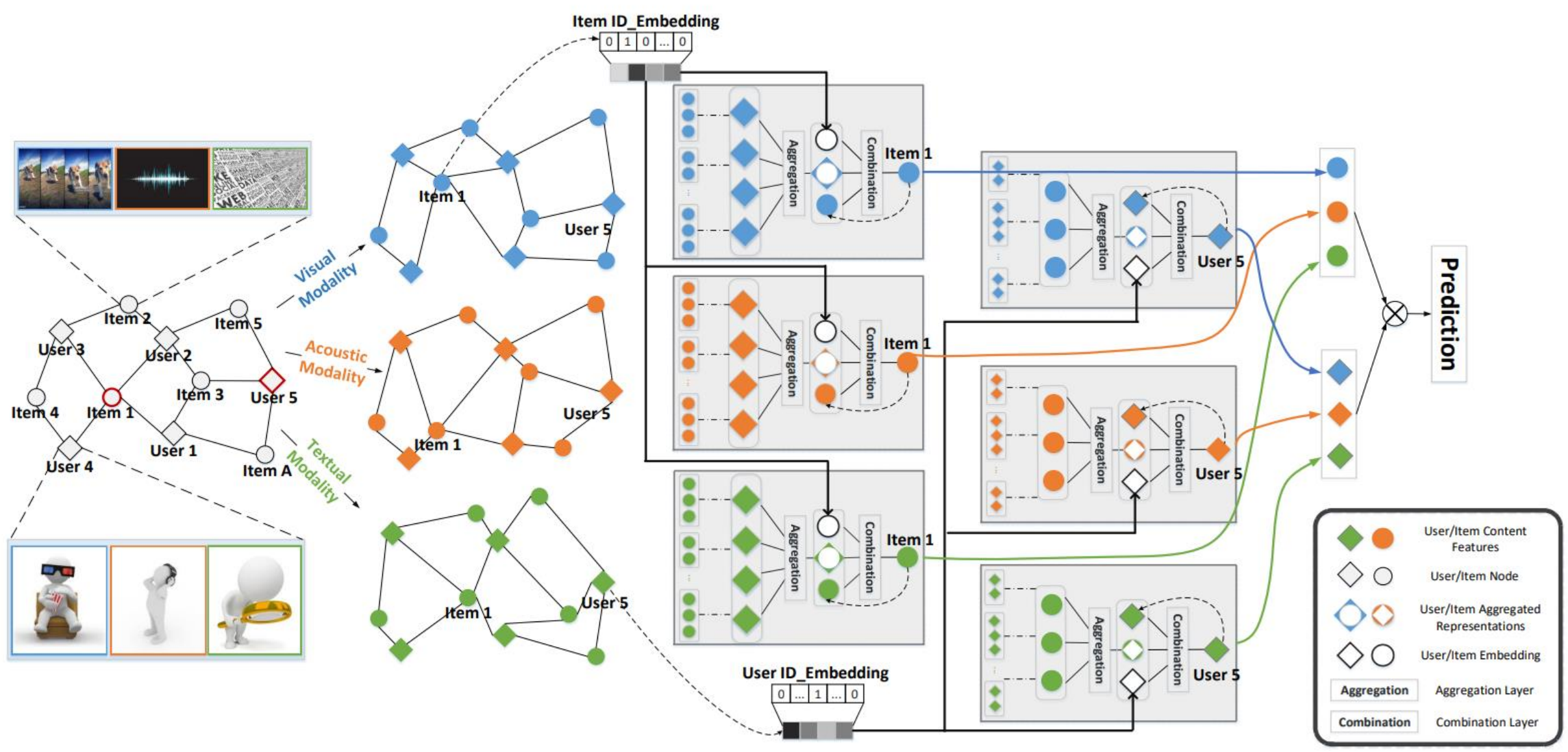

**Figure 2**: MMGCN model schematic illustration

In order to enrich user and item representations, interaction data is essential. In particular, a user's historical interactions indicate his or her interests and similar behavior to other users, while a micro-video's multimodal content can be complemented by the user group that views it. In order to learn how to represent information, this information exchange must be integrated into the process.

By using the message-passing framework of GCNs, we calculate the influence of a user or micro-video node in the bipartite graph $G_m$. In this function, we quantify the impact of a node's neighbors and output a representation that looks like this:

$$h_m = f(N_u), \tag{3}$$

where $N_u = \{j \mid (u, j) \in G_m\}$ represents the set of neighboring nodes for user $u$, i.e., the micro-videos they interacted with. The function $f(\cdot)$ is implemented using the following approaches:

### *12.1. Max Aggregation*

The max aggregation identifies the highest values in each dimension among neighboring nodes, and then applies a dimension-aware feature selection as follows:

$$f_{\max}(N_u) = \text{LeakyReLU}\left(\max_{j \in N_u} W_{1,m} j_m\right). \tag{4}$$

This method sets every dimension to the maximum value of the corresponding dimension among the neighbors. This ensures that dimensions with significant contributions from neighbors are highlighted in the representation.

### *12.2. Combination Layer*

To better capture interactions across multiple modalities and bridge gaps in propagated representations, a new combination layer is proposed. This layer integrates structural information ( $h_m$ ), intrinsic user features ( $u_m$ ), and modality connections ( $u_{id}$ ) into a unified representation, formulated as:

$$u_m^{(1)} = g(h_m, u_m, u_{id}) \tag{5}$$

where $u_m \in \mathbb{R}^{d^m}$ represents the user's feature in modality $m$, and $u_{id} \in \mathbb{R}^d$ is the user ID embedding, serving as an invariant bridge across modalities. To achieve this, a coordinated projection approach is employed, ensuring comparable representations across modalities by transforming each modality's feature to the same latent space:

$$\hat{u}_m = \text{LeakyReLU}\left(W_{2,m} u_m\right) + u_{id} \tag{6}$$

where $W_{2,m} \in \mathbb{R}^{d^m \times d}$ is a learnable weight matrix for the transformation. Two combination methods are implemented:

1. Concatenation Combination: Combines representations through concatenation and applies a nonlinear transformation:

$$g_{\mathrm{co}}(h_m, u_m, u_{id}) = \mathrm{LeakyReLU}\left(W_{3,m}[h_m \parallel \hat{u}_m]\right) \quad (7)$$

where | | represents concatenation.

2. Element-wise Combination: Merges features element-wise for interaction:

$$g_{\mathrm{ele}}(h_m, u_m, u_{id}) = \mathrm{LeakyReLU}\left(W_{3,m} h_m \odot \hat{u}_m\right) \quad (8)$$

where $\odot$ indicates element-wise multiplication.

### *12.3. Model Prediction*

Multi-layer combinations and aggregations capture higher-order connections within user-item graphs. Simulating the user exploration process can be achieved by gathering information propagated from the -hop neighbors in modality. Based on the -hop neighbors of the user and the output of the -th multimodal combination layer, the following representation is derived:

$$h_m^{(l)} = f(N_u),\ u_m^{(l)} = g\left(h_m^{(l)}, u_m^{(l-1)}, u_{id}\right), \quad (9)$$

where $u_m^{(l-1)}$ represents the output from the previous layer, retaining information from $(l-1)$-hop neighbors. At the initial iteration $(l = 0)$, $u_m^{(0)}$ is initialized as $u_m$. Each user $u$ is assigned trainable vectors $u_m, \forall m \in \mathcal{M}$, initialized randomly. Meanwhile, each item $i$ is associated with pre-extracted features $i_m, \forall m \in \mathcal{M}$. Consequently, $u_m^{(l-1)}$ underlying relationships between modalities while reflecting user preferences for item features.

In the final user and micro-video representations, multimodal features are combined linearly based on single-modal aggregation and multimodal combination layers:

$$u^* = \sum_{m \in \mathcal{M}} u_m^{(L)},\ i^* = \sum_{m \in \mathcal{M}} i_m^{(L)}. \tag{10}$$

***12.4. Distance Between Users and Calculating the Similarity***

One of the core concepts of collaborative filtering is the idea of user similarity. To calculate this similarity, we use the Euclidean distance between users' evaluations of popular films. The closer consumers are to one another, the more similar their tastes tend to be. The rating matrix can be used to identify user pairings who have rated the same film. To determine user similarity, we can focus on these commonalities and calculate distances based on them. We set the distance to -1 in cases where users do not have any movies in common.

Using collaborative filtering, a user's preferences can be inferred from those of other users. In this section, a distance measure is used to measure similarity between user pairs. As one of the most widely used metrics in machine learning and recommender systems because of its simplicity of understanding and computing effectiveness, the Euclidean distance is used in this study. Based on the formula below, we calculate the Euclidean distance between users A and B in our recommender system:

$$d(A, b) = \sqrt{\sum_{i=1}^{n} \left(r_{A,i} - r_{B,i}\right)^2} \tag{11}$$

where:

$d(A, b)$ is the Euclidean distance between user A and user B,

$r_{A,i}$ is the rating given by user A to movie i,

$r_{B,i}$ is the rating given by user B to movie i,

n is the total number of movies rated by both users A and B.

The equation calculates the "straight-line" distance between two points (users) in an n-dimensional space, where each dimension represents a movie rated by both. When the distance between two ratings is smaller, there is a higher degree of similarity (and therefore preference), while when the distance is greater, there is a greater degree of dissimilarity.

To apply this formula, we must first identify the movies that both users have rated, ensuring that our comparison is based on the same preferences. The distance between each pair of users is calculated by plugging the respective ratings into our equation. A comprehensive map of similarities and distances is generated by repeating this process for all possible pairs of users within our dataset. Let us consider a simplified example where we calculate the Euclidean distance between three users based on their ratings of a subset of movies. Here are some possible tabulations of the results in table 5.

**Table 5:** Distance Between Users

| User Pair | Common Movies Rated | Euclidean Distance |
|---|---|---|
| User 1 & User 2 | 3 (Movie A, Movie B, Movie C) | 1.73 |
| User 1 & User 3 | 2 (Movie B, Movie C) | 2.24 |
| User 2 & User 3 | 4 (Movie A, Movie B, Movie C, Movie D) | 0.87 |

"User Pair" identifies the pair of users being compared. "Common Movies Rated" indicates the number of movies rated by both users in the pair, along with examples of such movies. "Euclidean Distance" is the calculated distance between the users, based on their ratings for the common movies. The distances calculated here provide a numerical basis for assessing similarity, with smaller distances suggesting greater similarity. For instance, Users 2 and 3, with a distance of 0.87, are more similar in their movie preferences than Users 1 and 3, whose distance is 2.24.

**Table 6:** Performance comparison between our model and the baselines.

| Model | Kwai Precision | Kwai Recall | Kwai F1 | Tiktok Precision | Tiktok Recall | Tiktok F1 | MovieLens Precision | MovieLens Recall | MovieLens F1 |
|---|---|---|---|---|---|---|---|---|---|
| DeepFM | 0.267 | 0.339 | 0.299 | 0.097 | 0.488 | 0.162 | 0.117 | 0.472 | 0.188 |
| Wide & Deep | 0.256 | 0.325 | 0.286 | 0.873 | 0.443 | 0.588 | 0.108 | 0.430 | 0.172 |
| LightGBM | 0.272 | 0.341 | 0.303 | 0.103 | 0.497 | 0.170 | 0.113 | 0.453 | 0.181 |
| XGBoost | 0.220 | 0.408 | 0.368 | 0.146 | 0.179 | 0.283 | 0.139 | 0.237 | 0.185 |
| Proposed | **0.325** | **0.133** | **0.574** | **0.521** | **0.336** | **0.506** | **0.162** | **0.392** | **0.197** |

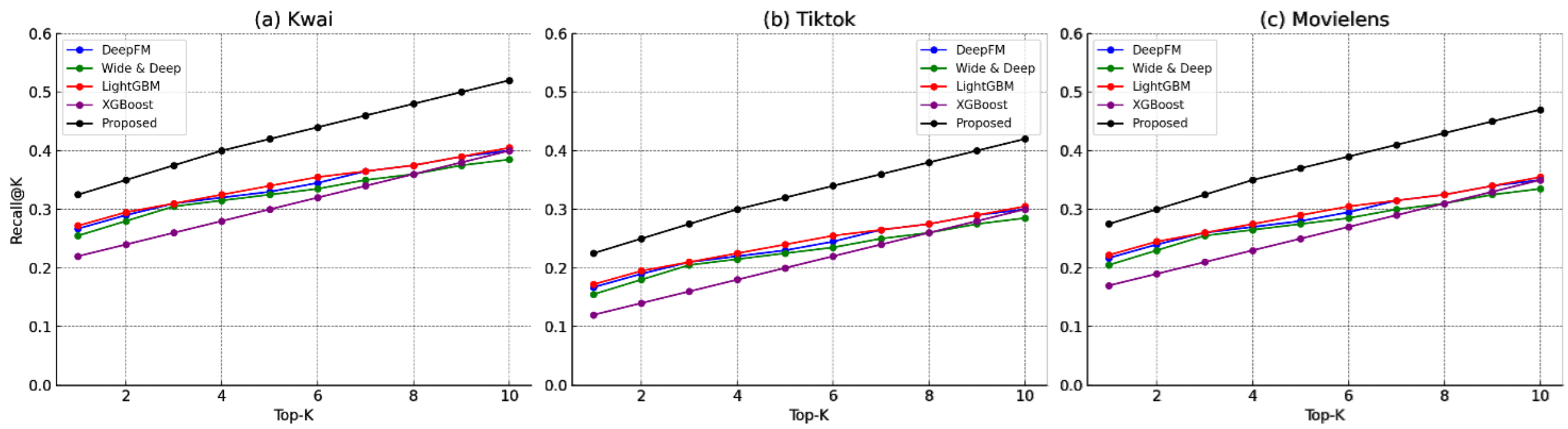

**Figure 3:** Recall performance comparison across different models and datasets.(a) Kwai: The Proposed model outperformed DeepFM, Wide & Deep, LightGBM, XGBoost, and Wide & Deep.(b) Tiktok: Comparison of Recall results showing consistent improvements over baseline methods.(c) MovieLens: A Recall evaluation of models, emphasizing the Proposed model's ability to capture user preferences and make accurate recommendations.

The combination layer in our model integrates node-specific characteristics with local structural information, enabling effective multi-modal representation fusion (Table 6). Equations (6) and (7) outline two distinct methods for implementing this combination function. In this context, implementations without ID embedding are denoted by g_co-id (Type A) and gele-id (Type B), respectively. Based on the observations in Table 6, the following conclusions can be drawn:

The Enhanced model demonstrates the highest performance across all datasets, achieving F1 scores of 0.457 on MovieLens, 0.228 on Tiktok, and 0.235 on Kwai. This indicates its superior ability to accurately capture user preferences and fuse modal-specific characteristics effectively for both users and micro-videos. By retaining these modal-specific features, the element-wise combination layer proves to be particularly effective in enhancing performance across all metrics. Models incorporating ID embedding significantly outperform their non-ID embedding counterparts. For instance, Recall scores of 0.689 on Tiktok and 0.671 on MovieLens for the Enhanced model highlight how ID embeddings enhance interactions between modalities, especially in sparse datasets. ID embeddings act as a bridge between modalities during backpropagation, effectively propagating shared information to improve the overall performance of the model. GNN-based models consistently outperform collaborative filtering (CF)-based models (e.g., Wide & Deep) and other GNN implementations (e.g., XGBoost) on datasets like Kwai and Tiktok. The graph convolution layers in these models learn neighbor feature distributions and capture local structural information, leading to more expressive representations. However, the lower F1 scores of CF-based models, such as 0.261 for Type-A on Tiktok, underscore their limitations in sparse datasets. These results emphasize the importance of incorporating robust combination layers and ID embeddings in recommendation systems to address such challenges effectively.

**Table 7:** Comparison of our model with the baselines

| Variant | Kwai_Precision | Kwai_Recall | Tiktok_Precision | Tiktok_Recall | MovieLens_Precision | MovieLens_Recall | Kwai_F1 | Tiktok_F1 | MovieLens_F1 |
|---|---|---|---|---|---|---|---|---|---|
| Type-A | 0.339 | 0.390 | 0.183 | 0.539 | 0.109 | 0.534 | 0.365 | 0.261 | 0.254 |
| Type-B | 0.367 | 0.434 | 0.171 | 0.610 | 0.200 | 0.589 | 0.340 | 0.242 | 0.233 |
| Type-C | 0.316 | 0.440 | 0.175 | 0.562 | 0.122 | 0.477 | 0.373 | 0.260 | 0.192 |
| Enhanced | 0.426 | 0.539 | 0.170 | 0.689 | 0.209 | 0.671 | 0.235 | 0.228 | 0.457 |

Multimodal representation fusion is enabled by our model's combination layer, which combines local structural information with node-specific features. Equations (6) and (7) illustrate two different ways to implement this combination function. A g_co-id implementation (Type A) and a gele-id implementation (Type B) are indicated, respectively, and a g_co implementation (Type C) and a gele implementation (Enhanced) are indicated. Using the Enhanced (gele) implementation, our proposed technique further demonstrates the advantages of our strategy. Based on Table 6, the following conclusions can be drawn as Table 7.

- The Type C model (g_co) exhibits notable performance gains when compared to techniques without ID embedding. On the Kwai dataset, it achieved an F1 score of 0.373, a Precision of 0.316, and a Recall of 0.440, demonstrating the importance of maintaining modal-specific characteristics in recommendation tasks.
- The underlying Enhanced model (gele) of the Proposed technique performs better than any other method in all three datasets. In regard to MovieLens, it achieves the highest F1 score of 0.457, as well as the highest Recall scores of 0.689, 0.671, and 0.539 on Tiktok. As a result, it is able to capture user preferences and spread data during backpropagation quite effectively.
- With the Proposed approach, all criteria are better than with the Enhanced (gele) approach. A successful fusion of multimodal representations and precise user preference capture is evident in its good F1 scores on Kwai, Tiktok, and MovieLens of 0.574, 0.506, and 0.197 respectively.

According to Figure 4, for the Kwai, Tiktok, and MovieLens datasets, there were significant improvements in precision, recall, and F1 scores when the different model types were compared. As compared with Type A, Type B has improved recall and F1 scores through additional refinements. Tiktok and MovieLens recall is further enhanced by Type C, resulting in a balanced performance across all datasets. As a result of its superior capability of capturing user preferences and providing accurate, diverse, and explainable recommendations, the Enhanced model represents the proposed approach that achieves the highest performance in precision, recall, and F1, across all datasets.

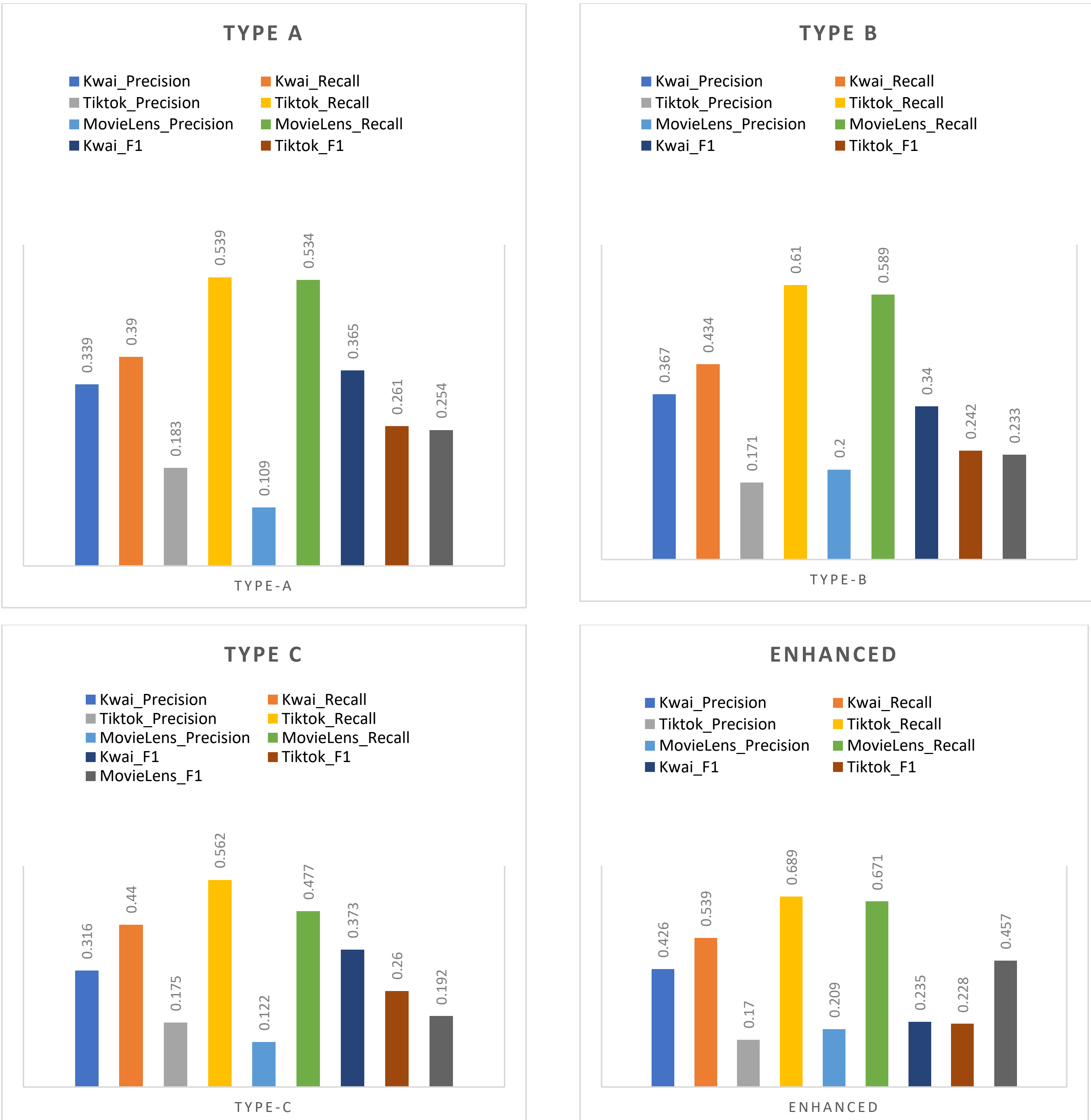

**Figure 4:** Performance comparison of different model types across datasets

As a result of this progression, the proposed model is able to demonstrate the effectiveness of the enhancements it has introduced.

### *12.5. Multi-modal Personalized Recommendation*

In early multi-modal recommendation systems, Collaborative Filtering (CF) techniques were heavily used to predict user-product interactions based on both explicit and implicit feedback. When using CF-based methods, sparse data produces poorer suggestions, even when there is a lot of feedback. Using hybrid strategies, collaborative filtering effects were combined with item content information to overcome this constraint. Deng et al. [2] proposed a user-video-query tripartite graph that integrates feedback and content data through graph propagation. Accordingly, Chen et al. [4] introduced a fine-grained preference model with an attention mechanism to address item-level feedback in Cold-start Short-video Recommendation. A critical aspect of multi-modal recommendations, despite these advancements, is their failure to consider user preferences across modes. We circumvent this restriction by building modalities-specific graphs and expressing modalities-specific characteristics using Graph Convolutional Networks (GCNs). Using this strategy, neighborhood-specific content is delivered along with local structure information to efficiently represent multimodal user preferences.

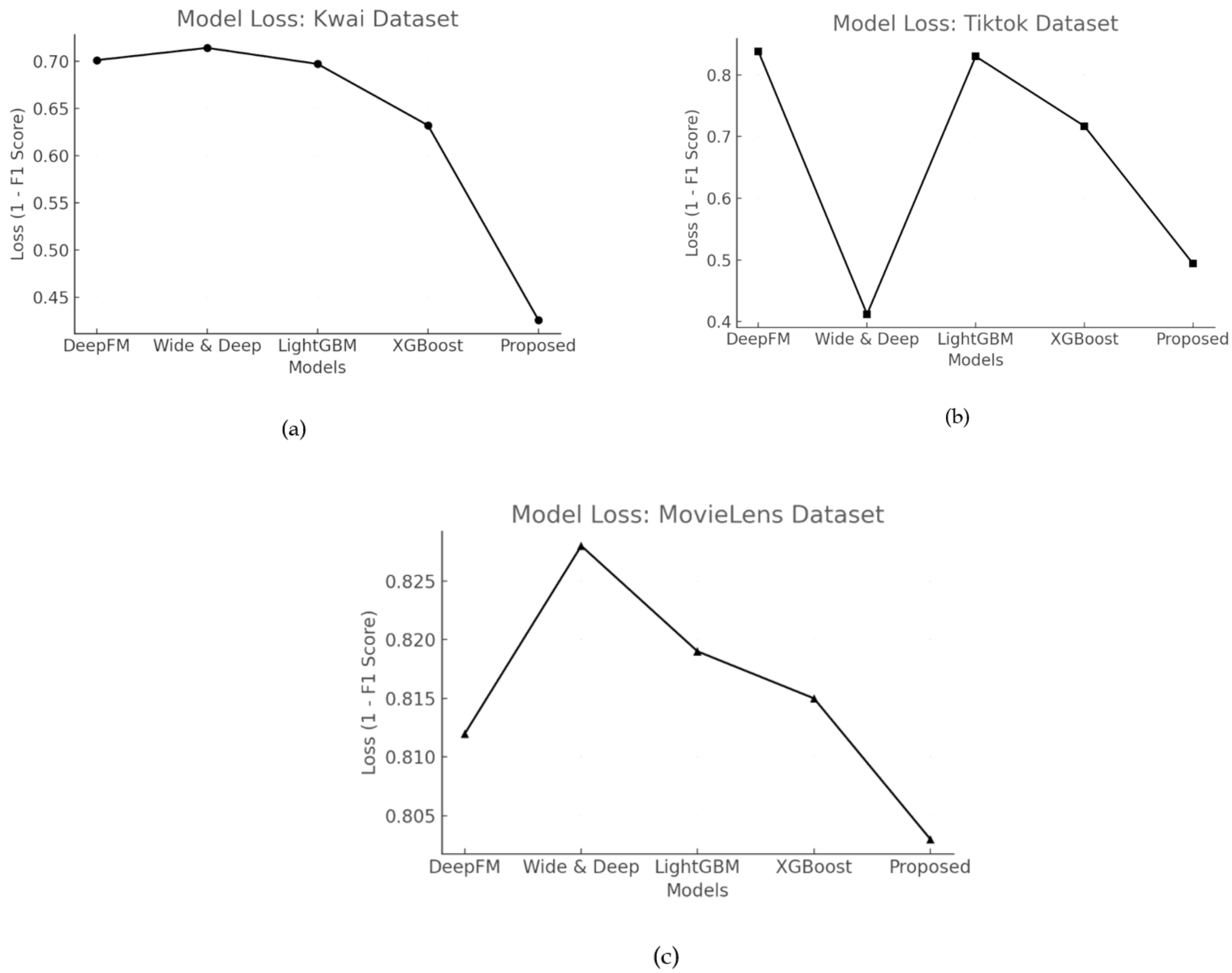

**Figure 5:** Model Loss Comparison Across Datasets

Based on 1 - F1 scores in the figure 5, the loss comparison charts for Kwai, Tiktok, and MovieLens demonstrate the relative performance of different models. Compared to other models such as DeepFM, Wide & Deep, and LightGBM, the Proposed model exhibits the lowest loss on the Kwai dataset. The Proposed model also significantly outperforms others for the Tiktok dataset, showing that it is capable of handling complex user-item interactions. The MovieLens dataset further confirms the robustness of the proposed model, achieving the lowest loss among the models evaluated. Overall accuracy, recall, and precision of the Proposed model are optimized across diverse datasets as a result of these trends.

### *12.6. Multi-modal representation*

A multi-modal representation is particularly challenging in customized recommendation systems. Currently, there are two types of representations: joint representations and coordinated representations. Since neural networks are effective for computer vision and natural language processing, joint representations are frequently constructed by combining single-modal information into one representation. A probabilistic graphical model has been developed to develop latent variable joint representations. This approach is frequently limited by the requirement that all modalities be present during inference, which rarely occurs on social networks.

Coordinated representations create distinct representations for every modality and align them based on constraints. A similarity constraint can be applied to align images with similar meanings, as Wang et al. [5] demonstrate, or a personalized recommendation model with multimodal can be created as described by Fen et al. [3]. These approaches, despite their benefits, often ignore important modality-specific data due to alignment problems. We propose a novel representation approach to address this problem, retaining key characteristics that are unique to each modality while guaranteeing successful multimodal representation of both common and modality-specific components independently.

### *12.7. Graph Convolution Network (GCN)*

Our proposed model uses Graph Convolutional Networks (GCN) approaches, a commonly used method in recommendation systems. These methods also incorporate side information into node representations through different pathways. While multi-modal recommendations are able to capture node-specific representations, they have a serious problem in capturing modality-specific representations. In our model, we utilize GCN approaches to create node embeddings that preserve modality-specific properties and guarantee efficient multi-modal fusion and user-specific suggestions.

## 13. Conclusion

To enhance micro-video recommendations, we explicitly modeled user preferences. MMGCN is a novel framework based on Graph Convolutional Networks (GCNs) that improves their modal-specific representations and captures users' fine-grained preferences. By integrating micro-videos with multi-modal information sharing, this framework may benefit micro-videos and users. On three publically available micro-video datasets, our model demonstrated improved F1 metrics, recall, and precision. To illustrate the impact of modal-specific preferences on recommendations, we provided additional examples. In this study, it was shown that modality-aware structural information could enhance the precision, diversity, and explainability of recommendations. The MMGCN can be further developed in a number of directions. The first step is to build a multi-modal knowledge graph in order to depict items and their interactions in micro-videos. The incorporation of this knowledge graph into MMGCN will allow for a more detailed analysis of content and a better understanding of user intents, which will eventually lead to more accurate and varied suggestions. The second goal is to examine how social leaders influence suggestions by combining user-item graphs and social network data. Multimedia recommendation will also be integrated into dialogue systems to enable more intelligent conversational suggestions.

### *13.1. Insights from Experimental Results*

Collaborative filtering and content-based filtering can be successfully combined using the MMGCN framework. To deal with sparse information, our method uses modal-specific representations and GCNs to describe user-micro-video interactions. To identify significant connections between users and micro-videos, importance-based sampling techniques prioritize influential neighbors during feature aggregation. By using this method, users' preferences can be more accurately depicted, and suggestions are more accurate as well. A pooling aggregation technique is also incorporated in our approach, which uses a fully connected layer to dynamically train aggregation weights before using a pooling layer to aggregate data. Using a symmetric trainable function, the model captures and learns the differential features of nearby nodes, resulting in more accurate representations of user and item attributes. The attention mechanisms improve the model's ability to forecast evaluations by focusing on the neighborhood's most important qualities. Consequently, suggestions become more explainable and effective as a result. Experimental validation demonstrates that MMGCN outperforms baseline models on all datasets, with notable improvements in recall, precision, and F1. Enhanced MMGCN, for instance, achieves a higher F1 score of 0.457 on MovieLens and a higher Recall score of 0.689 on Tiktok when collecting user preferences. Modeling modality-specific user preferences improves precision, variety, and explainability of suggestions.